\documentclass[onecolumn,authoryear]{els-mrw} 

\usepackage{aas_macros}
\usepackage{amsmath,amssymb,amsfonts,amsthm,makeidx,graphicx}
\usepackage{txfonts}
\usepackage{helvet}
\usepackage{soul}
\usepackage{subcaption}
\usepackage{xcolor}

\newcommand{\eqn}[1]{equation~(\ref{#1})}
\newcommand{\eqns}[2]{equations~(\ref{#1}) and (\ref{#2})}

\newcommand{\secn}[1]{Section~\ref{#1}}
\newcommand{\fig}[1]{Figure~\ref{#1}}

\newcommand{\angstrom}{\text{\normalfont\AA}}

\newcommand{\Msun}{\rm{M_{\odot}}}
\newcommand{\LUV}{ L_{\rm UV} }

\newcommand{\fesc}{f_{\rm esc} }
\newcommand{\QHII}{Q_{\rm HII}}
\newcommand{\xHI}{x_{\rm HI}}
\newcommand{\fcoll}{f_{\rm coll}}

\newcommand{\xHIbar}{\bar{x}_{\rm HI}}

\begin{document}

\chapter{Reionization and its sources}\label{chap1}

\author[1]{Anirban Chakraborty}%
\author[1]{Tirthankar Roy Choudhury}%
\address[1]{\orgname{National Centre for Radio Astrophysics, Tata Institute of Fundamental Research}, \orgaddress{Pune University Campus, Ganeshkhind, Pune 411007, India}}

%%\address[2]{\orgname{Name of Institute}, \orgdiv{Division or Department}, \orgaddress{Address of Institute}}%%

\articletag{Chapter Article tagline: update of previous edition,, reprint..}

\maketitle
\begin{abstract}[Abstract]

Reionization represents an important phase in the history of our Universe when ultraviolet radiation from the first luminous sources, primarily stars and accreting black holes, ionized the neutral hydrogen atoms in the intergalactic medium (IGM). This process follows the ``Dark Ages'', a period with no luminous sources, and is initiated by the formation of the first sources, marking the ``Cosmic Dawn''. Reionization proceeds through multiple stages: initially, ionized bubbles form around galaxies, then expand and overlap across the IGM, culminating in a fully ionized state, with neutral hydrogen remaining only in dense regions. Understanding reionization involves a diverse range of physical concepts, from large-scale structure formation and star formation to radiation propagation through the IGM. Observationally, reionization can be explored using the cosmic microwave background (CMB), Lyman-$\alpha$ absorption, high-redshift galaxy surveys, and emerging 21~cm experiments, which together offer invaluable insights into this transformative epoch.

%Approx. 100-150 words abstract of the chapter, used to summarise the work. 
\end{abstract}

\begin{keywords}[Keywords ]
Reionization, Intergalactic Medium, First Stars, Primordial Galaxies, Early Universe, Cosmology, Large-scale structure of the universe, Lyman alpha forest, Cosmic microwave background radiation

%5-10 words that embody the key topics in the chapter. What terms would someone put into a search engine if they were looking for a chapter like this?(\url{https://astrothesaurus.org/}, please use this site for the keywords to be included inthe chapter)
\end{keywords}
\begin{glossary}[Nomenclature]
\begin{tabular}{@{}lp{34pc}@{}}
CMB & Cosmic Microwave Background \\
DM & Dark Matter \\
CDM & Cold Dark Matter \\
IGM & Intergalactic medium \\
UV & Ultraviolet \\
EoR & Epoch of Reionization \\
IMF & Initial Mass Function \\
ISM & Interstellar medium \\
CGM & Circumgalactic medium \\
AGN & Active Galactic Nuclei \\
kSZ & kinetic Sunyaev-Zel’dovich \\
GP & Gunn-Peterson \\
HST & Hubble Space Telescope \\
JWST & James Webb Space Telescope \\
LAEs & Lyman Alpha Emitters \\
LBGs & Lyman Break Galaxies \\
SKA & Square Kilometre Array \\
DM & Dispersion Measure \\
FRBs & Fast Radio Bursts \\

\end{tabular}
\end{glossary}

\section{Learning Objectives}\label{chap1:sec1}

\begin{BoxTypeA}[chap1:box1]{}

By the end of this chapter, the reader is expected to understand:
\begin{itemize}
    \item The meaning of cosmic reionization and its significance in the evolution of the Universe.
    \item How reionization proceeds, both qualitatively and quantitatively, with time.
    \item The different kinds of ionizing sources that may have been responsible for driving reionization.
    \item How reionization can be studied through various observations such as the cosmic microwave background (CMB), Lyman-$\alpha$ absorption, high-redshift galaxy surveys, and upcoming 21~cm experiments. 
\end{itemize}

\end{BoxTypeA}

\section{Introduction}\label{chap1:sec2}

Our understanding of the Universe's evolutionary history is based on the foundations of the standard hot Big Bang model, which has received considerable observational support over time. According to this paradigm, the Universe was extremely hot and dense in its initial moments, with photons existing in thermal equilibrium with matter. As the Universe expanded, its temperature and density gradually dropped. Around three minutes after the Big Bang, nuclei of elements heavier than hydrogen (mainly helium and lithium) formed through a process called nucleosynthesis. About 380,000 years later, the temperature decreased further, allowing hydrogen, helium, and lithium nuclei to capture electrons and form neutral atoms. With the formation of the first atoms, photons were no longer scattered by free electrons in the plasma and could travel freely through the Universe. These photons are observed today as the cosmic microwave background (CMB) radiation, providing the earliest possible view of the Universe. Soon after the recombination epoch, the Universe entered a phase called the ``Dark Ages", where no significant radiation sources existed, and hydrogen remained largely neutral. During this period, small perturbations in the matter density present at the time of recombination began to grow due to gravitational instability. In this process, slightly overdense regions accreted matter from surrounding underdense regions due to their self-gravity, becoming increasingly overdense. As these overdense regions continued to accrete more and more matter, they expanded at an increasingly slower pace compared to the background Universe, ultimately breaking away from the background expansion and collapsing under their own gravity. During this gravitational collapse, the infalling matter particles interacted gravitationally amongst themselves and exchanged energies, eventually reaching a state of virial equilibrium. This marked the formation of gravitationally bound structures known as ``dark matter haloes", as the matter component of the Universe is dominated by the dark matter (DM). In the context of the cold dark matter (CDM) model of structure formation, these newly formed DM haloes continue to evolve, growing in mass and size either by accreting material from their surroundings or by merging with other haloes.

\begin{figure}[t]
\centering
\includegraphics[scale=0.5]{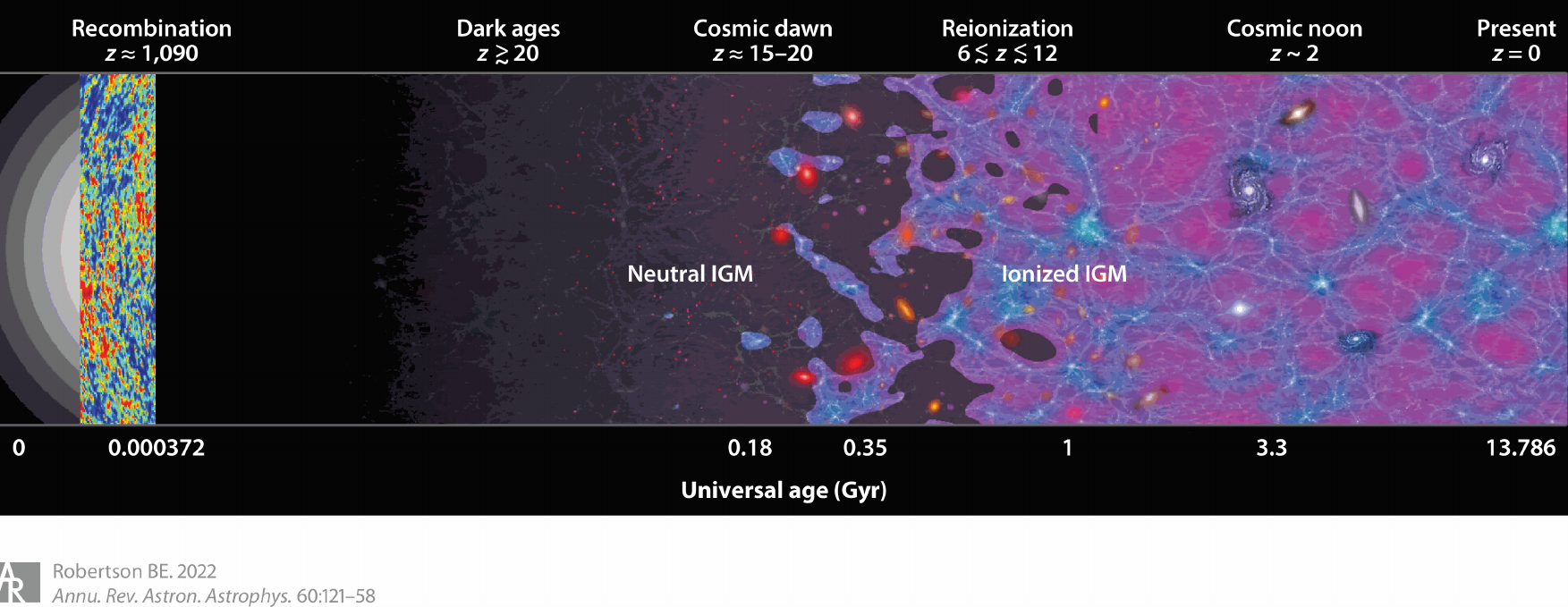}
\caption{An illustration highlighting the various milestones in the evolutionary history of our Universe, since the Big Bang. The figure shows the evolution of the ionization state of the IGM due to ionizing radiation from the earliest generation of luminous sources, alongside the corresponding evolution in their abundances. Image Credits: \cite{Robertson2022} }
\label{fig:cosmic_timeline}
\end{figure}

These collapsed dark matter-dominated haloes form potential wells, into which the baryons passively fall. It should, however, be kept in mind that most of the baryons at high redshifts do not reside within these collapsed DM haloes, they rather reside as diffuse gas within the intergalactic space which is known as the intergalactic medium (IGM). The baryonic gas, when attracted into the dark matter potential well, acquires kinetic energy and is heated up to the virial temperature of the halo via shocks. Further evolution can only take place if this shock-heated gas can somehow cool. If the mass of the halo is sufficiently high (i.e., the potential well is deep enough), the gas will be able to dissipate its internal energy via atomic or molecular transitions and fragment within the halo. This produces conditions appropriate for gas to condense and form stars (and eventually, galaxies), thereby marking the end of the era of the ``Dark Ages" and the onset of ``Cosmic Dawn''. Once these earliest luminous sources (i.e., stars inside galaxies) form, they produce copious amounts of ultraviolet (UV) radiation through thermonuclear reactions in their cores. In addition to galaxies, a population of early black holes may have also contributed to this UV radiation. Some of this radiation escapes out of these systems, carrying photons with energies of $\geq 13.6~\text{eV}$ that ionize the surrounding neutral hydrogen atoms, initiating the ``Epoch of Reionization" (EoR). As more luminous sources formed, the individual ionized bubbles localized around the earliest sources began to overlap, gradually expanding to fill larger volumes of the Universe with ionized gas. This process finally concluded around a billion years after the Big Bang, marking a significant phase transition in the ionization state of hydrogen atoms in the IGM. By the end of reionization, the hydrogen in the Universe was almost fully ionized (see \fig{fig:cosmic_timeline} for an illustration of the timeline). The sources, primarily stars, that ionize hydrogen are also capable of singly ionizing helium (HeI $\rightarrow$ HeII) through photons with energies of $\geq 24.6~\text{eV}$ and the first reionization of helium is thus believed to be roughly contemporaneous with that of hydrogen. However, the removal of the second electron from helium (known as HeII reionization) requires photons with even higher energies ($\geq 54.4~\text{eV}$), which are generated by more powerful sources, such as accreting black holes (commonly referred to as quasars), and occurs at a much later time. In this article, we will concentrate mostly on hydrogen reionization by the first luminous sources.

From an observational standpoint, the era of cosmic reionization is a largely unexplored phase of the Universe. It serves as a crucial link between the early Universe, which we study through the CMB, and the late Universe, which is well understood through observations of galaxies, clusters, quasars, and other astronomical sources.  Over the last few years, the availability of high-quality datasets obtained from various cutting-edge telescopes that probe the earliest epochs of our Universe has provided a fresh impetus to the study of cosmic reionization and the first stars. Moreover, understanding cosmic reionization is essential for studying how structures in the Universe formed and evolved since it not only results from the formation of the first luminous sources but also influences the formation and evolution of structures at later epochs. For instance, during reionization, ionizing photons not only ionized neutral atoms but also heated the IGM through transfer of the excess energy carried by the free electrons. This additional heating in ionized regions affects the subsequent formation of stars by either expelling gas or preventing gas from being efficiently cooled, particularly in low-mass haloes. The details of the reionization process, among other factors, depend on the properties of the ionizing sources. Therefore, reionization studies provide a complementary approach to understanding the earliest generation of luminous sources through their impact on the IGM — particularly those too faint to be detected directly, even with next-generation telescopes.

In this chapter, we present an overview of our current understanding of cosmic reionization and its probable sources and highlight how the Epoch of Reionization can be probed through a wide variety of observations. The chapter is organized as follows:  In \secn{chap1:sec3}, we briefly describe the stages of reionization and how the same can be studied using theoretical models. This is followed by a discussion of the potential sources of ionizing radiation that may have contributed to cosmic reionization in \secn{chap1:sec4}. We discuss the different observational probes that can shed light on the properties of the ionizing sources and the missing details about reionization in \secn{chap1:sec5}. 

\section{The stages of hydrogen reionization}\label{chap1:sec3}

\begin{figure}[t]
\centering
\includegraphics[scale=0.5]{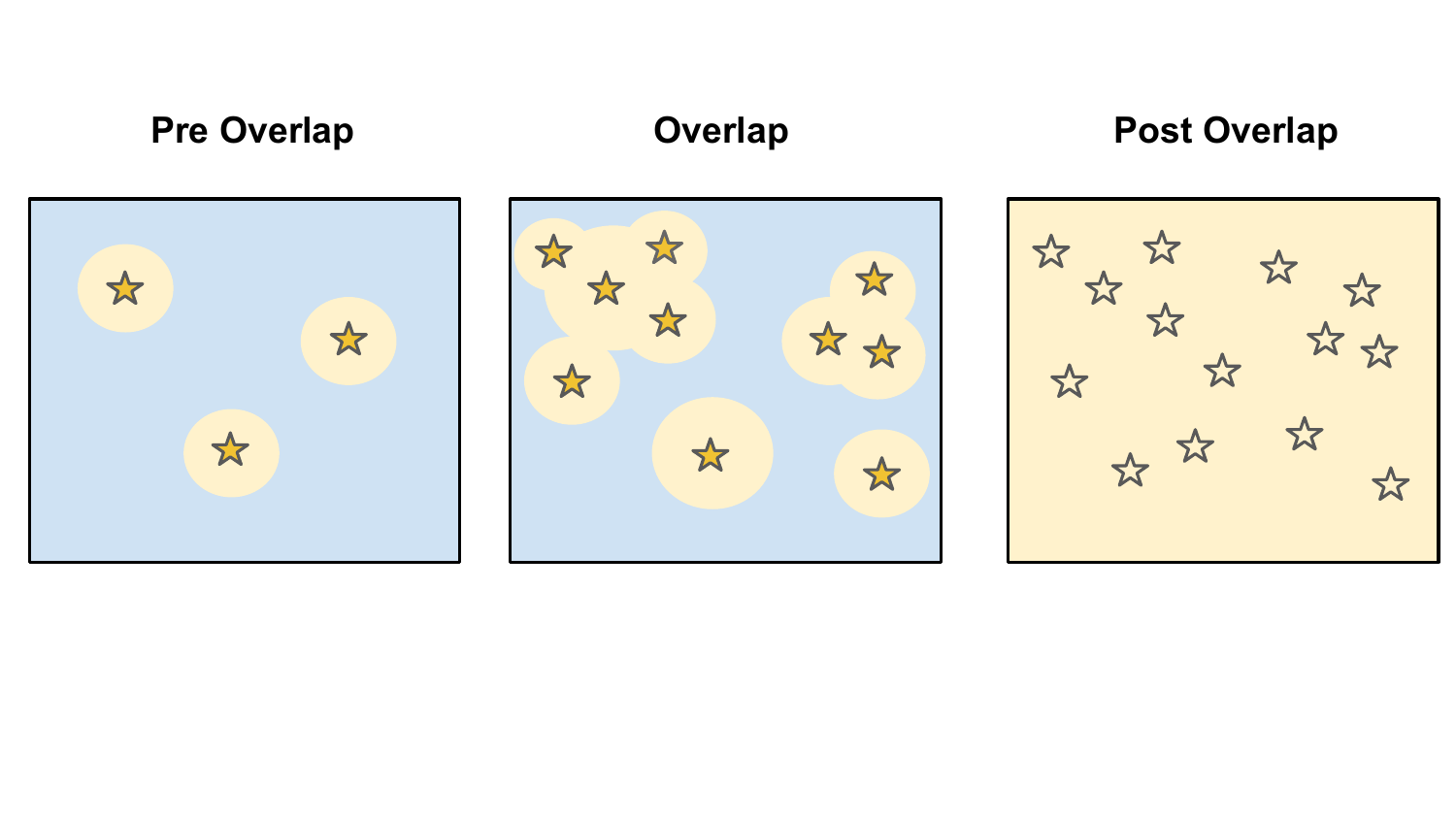}
\caption{A cartoon illustration depicting the different phases of reionization. Initially, ionized bubbles form around individual sources (represented by star symbols here), then expand and merge with other surrounding bubbles until the entire IGM is completely ionized, with neutral hydrogen remaining only in dense regions (not shown in the figure).}
\label{fig:reionization_stages}
\end{figure}

The reionization of hydrogen occurs in several stages, as shown in \fig{fig:reionization_stages}, starting with the ``pre-overlap" phase. In this initial stage, individual ionizing sources light up and begin ionizing the surrounding medium, which is predominantly composed of hydrogen. This process results in the formation of a region of photoionized gas, known as an HII region, around each source. These cosmological HII regions gradually grow as ionizing radiation propagates deeper into the neutral intergalactic medium. Since the first ionizing sources are thought to form in high-density regions of the Universe, their ionizing fronts must navigate through these dense environments, where the recombination rates are also typically high. Beyond these dense regions, however, the ionizing fronts spread more easily through the low-density voids of the Universe. As these HII regions expand, they leave behind some clumps of neutral, high-density gas, which, owing to their higher recombination rates, require more ionizing photons to become ionized than are available at this stage. During this time, the IGM exists in the form of a two-phase medium with the highly ionized (HII) regions separated from neutral regions by ionization fronts. Although the intensity of ionizing radiation is initially low, it gradually increases over time during the pre-overlap stage.

Because the first collapsed objects tend to form in the densest regions of the universe, their distribution is not a direct tracer of the overall matter distribution but is instead amplified in the highest-density peaks. As a result, the earliest luminous sources are strongly clustered. Consequently, reionization quickly progresses into its second stage, known as the ``overlap" phase. During this phase, neighboring HII regions begin to coalesce, and the ionizing background rapidly increases as an increasing number of newly formed sources power each HII region. Therefore, the ionization fronts can now expand into even regions with high gas density that had previously remained neutral. By the end of this phase, all low-density regions of the IGM as well as some of the high-density regions that were located close to the ionizing sources are completely ionized. A small fraction of neutral gas however remains confined within dense structures that were located far from any ionizing source. As more and more ionizing sources continue to form, these high-density regions are also expected to get gradually ionized, leading to an increasingly uniform ionizing background. This final stage, known as the ``post-overlap" phase, continues indefinitely - in fact, collapsed objects such as galaxies still have neutral gas trapped inside them in the local Universe.

From the above discussion, it is evident that cosmic reionization progresses at different rates in different regions of the Universe. On average, regions with a higher concentration of ionizing sources undergo reionization more rapidly compared to those where such sources are scarce. This overall description of reionization, where the process begins in high-density regions hosting ionizing sources and extends outward to low-density voids, is commonly referred to as ``inside-out'' reionization. While this holds true during the initial stages of reionization, there is a shift in the pattern of reionization when it approaches the later stages of the overlap phase. In its final stages, the high recombination rates in dense regions dominate the process and reionization penetrates deep into underdense regions before gradually evaporating the denser neutral regions. This results in a more ``outside-in'' progression, with low-density regions getting ionized earlier than the high-density regions. In this regard, reionization can be viewed as a process that essentially evolves in an ``inside-out-in'' manner. Understanding the various stages of reionization involves solving complex, non-linear physics, which is feasible only through detailed radiative transfer simulations. These simulations aim to model the transport of radiation through the cosmological matter field while simultaneously tracking the thermal and ionization states of the gas. Given the broad range of spatial and temporal scales involved, such studies are computationally intensive \citep{Gnedin2022}. Recently, efforts have been made to study reionization using approximate yet computationally efficient simulations known as semi-numerical simulations. These simulations replace detailed radiative transfer with a simplified photon-counting algorithm and typically employ coarser resolutions than full simulations \citep{Furlanetto2004, Mesinger2007, Choudhury2009, Santos2010, Mesinger2011, Choudhury2018, Hutter2018, Mondal2021_ReionYuga, Schaeffer2023}.

The progress of reionization can be approximated using an evolution equation for the globally averaged ionization fraction \citep{Madau1999}
\begin{equation}
    \frac{dQ_{\rm HII}(t)}{dt}=\frac{1}{\bar{n}_{H}} \frac{dn_{\mathrm{ion}}(t)}{dt}- Q_{\rm HII}(t)~\chi_\mathrm{He}(t)~\bar{n}_{\rm H}~a^{-3}(t)~\mathcal{C}(t)~\alpha_R(T),
\label{eqn:Madau_Eqn}
\end{equation}
where $\QHII$ is the fraction of volume ionized, 
$dn_{\mathrm{ion}}(t)/dt $ denotes the number density of ionizing photons produced by the sources and available for reionization in the IGM per unit time per unit comoving volume, $\bar{n}_H$ is the present-day mean comoving number density of hydrogen, $a(t)$ is the cosmological scale factor, $\alpha_R(T)$ is the recombination rate coefficient which depends on the temperature $T$ of the medium, $\chi_{\mathrm{He}}(t)$  denotes the number of free electrons per hydrogen atom (which includes the excess contribution from helium, depending on its ionization state) and $\mathcal{C}(t)$ is the clumping factor that accounts for enhanced recombinations in an inhomogeneously ionized IGM. The above equation can be interpreted as follows: the total number of ionizing photons produced (the first term on the right-hand side) is used either to ionize regions that were previously neutral (the term on the left-hand side) or to re-ionize recombined atoms within already ionized regions (the second term on the right-hand side). This equation can be readily solved to obtain the reionization history $Q_{\rm HII}(t)$ when supplemented with a model for the emissivity of ionizing
photons $dn_{\mathrm{ion}}(t)/dt$ and clumping $\mathcal{C}(t)$. Despite the simplifying assumptions in the above equation (e.g., ignoring density dependence in the growth of ionized regions and assuming ionizing photons are almost immediately absorbed by the diffuse IGM), it is widely used to model the progression and history of reionization, and also fares well when compared with more detailed radiative transfer simulations.

Before proceeding further, let us look at the reionization histories for a simplistic model where the comoving number density of ionizing photons available for reionization is parameterized as $n_{\mathrm{ion}}(t) = \bar{n}_H ~\zeta(t)~\fcoll(M_{\rm min}, t)$. Here, $\zeta(t)$ denotes the number of ionizing photons injected into the IGM per hydrogen atom inside collapsed objects, while $\fcoll(M_{\rm min},t)$ denotes the fraction of mass contained in collapsed objects of mass $M \geq M_{\rm min}$ at a given cosmic epoch $t$. In general, the ionization efficiency parameter $\zeta$ encapsulates information about several other poorly understood astrophysical processes such as the star-forming efficiency within the collapsed structure, the number of ionizing photons produced per unit stellar mass, and the escape probability of these photons. For simplicity, we assume that the two unknown parameters $\zeta$ and $\mathcal{C}$ appearing in \eqn{eqn:Madau_Eqn} do not evolve with time. Therefore, under these assumptions, we can rewrite \eqn{eqn:Madau_Eqn} as
\begin{equation}
    \frac{dQ_{\rm HII}(t)}{dt}=\zeta~\dfrac{d\fcoll(M_{\rm min}, t)}{dt} - Q_{\rm HII}(t)~\chi_\mathrm{He}(t)~\bar{n}_{\rm H}~a^{-3}(t)~\mathcal{C}~\alpha_R(T),
\label{eqn:reion_eqn}
\end{equation}

\begin{figure}[t]
\centering
\includegraphics[scale=0.45]{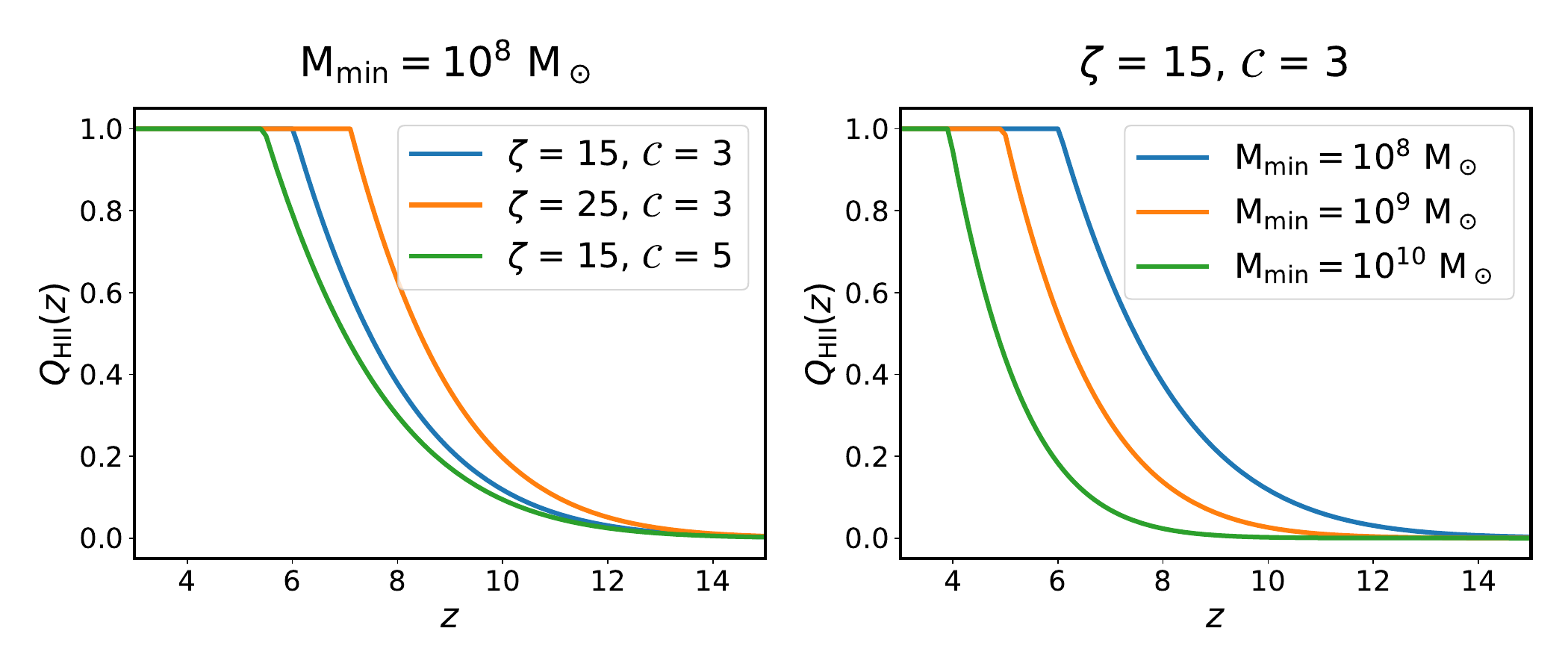}
\caption{The evolution of the ionized volume fraction $Q_{\rm HII}$ of the Universe as a function of redshift $z$ corresponding to some representative values of the ionization efficiency parameter $\zeta$ and clumping factor $\mathcal{C}$ (\textit{left-hand panel}) and the minimum mass $M_{\rm min}$ of haloes capable of producing ionizing photons  (\textit{right-hand panel}). It has been assumed that none of these three parameters vary over time.}
\label{fig:reionization_history}
\end{figure}

The left-hand panel of \fig{fig:reionization_history} shows the reionization histories obtained by numerically solving \eqn{eqn:reion_eqn} for representative values of $\zeta$ and $\mathcal{C}$. Here, we fix the minimum mass of haloes $M_{\rm min}$ contributing to the ionizing photon budget to be 10$^8~\Msun$, appropriate for atomically cooled haloes (to be discussed in the next section).  From the figure, one can see that when the efficiency of ionizing photon production is increased (i.e., the value of $\zeta$ becomes larger, keeping $\mathcal{C}$ unchanged), reionization progresses faster and completes earlier due to the availability of a higher number of ionizing photons. On the other hand, if the IGM is assumed to be excessively clumpy (i.e., the value of $\mathcal{C}$ becomes larger, keeping $\zeta$ unchanged), the progress of reionization is slowed down since the recombination rate gets enhanced due to increased clumpiness and hence more ionizing photons are now consumed in ionizing the recombined atoms. We show the effect of varying the value of  $M_{\rm min}$ (keeping $\zeta$ and $\mathcal{C}$ unchanged) in the right-hand panel of \fig{fig:reionization_history}. Raising the value of $M_{\rm min}$ leads to a delayed onset of reionization, since larger haloes form comparatively later in the Universe due to the hierarchical nature of structure formation.

\section{Sources responsible for cosmic reionization}\label{chap1:sec4}

\subsection{Star-forming galaxies}

The stars in the earliest galaxies are considered to be the most promising sources of ionizing UV photons, largely due to their high abundance during this era. Understanding the key factors that influence the formation of stars and the production of ionizing radiation from these systems at such early epochs is therefore of great importance. As mentioned earlier, these stars form within dark matter haloes massive enough to attract the surrounding baryonic material (mainly, gas). However, for this accreted gas to be able to form stars, it must first undergo rapid gravitational collapse — i.e., the force of gravity must exceed its internal pressure. But, this alone is not sufficient, since the temperature of the gas rises rapidly during contraction, leading to an increase in its internal pressure that resists further gravitational collapse. Eventually, the gas, after being shock-heated to temperatures close to the halo's virial temperature, settles into an equilibrium configuration where it is supported against gravity by its thermal pressure. Without any other dissipative mechanisms, the gas would remain in this low-density, hot state indefinitely, preventing star formation. Therefore, a second key requirement is that the gas must cool and condense into dense molecular clouds, where star formation can finally occur.

In principle, gaseous matter can cool down by losing its internal energy through various atomic processes, such as collisionally-excited emission and different types of continuum radiation, including bremsstrahlung, recombination, and collisional ionization. Such cooling is effective only when it takes place much faster than the free-fall timescale of the system. However, at high redshifts, where the gas is primarily composed of hydrogen and helium atoms, the dominant cooling process is the collisional excitation of atoms and subsequent radiative de-excitation. In this case, the atoms get excited to a higher energy level because of a collision with another atom and then return to the lower state by emitting photons. If these photons are able to escape the halo, carrying away energy, the system will cool down. However, below a temperature of 10$^4$ K (corresponding to a halo mass $M \sim 10^8$ M$_\odot$), cooling via atomic transitions is not very effective because collisions among the atoms do not have sufficient energy to excite the atoms. In the absence of heavy elements, cooling at low temperatures ($T \leq 10^4$ K) is primarily facilitated by molecules like hydrogen (H$_2$), whose rotational and vibrational levels can be excited by low-energy electrons. In these pristine conditions, the formation of H$_2$, catalyzed by free electrons left over from the epoch of recombination, enables efficient cooling even in small haloes (commonly known as minihaloes) with masses around $M \sim$ 10$^{6}$ M$_\odot$.

Tracking the precise cooling mechanisms, followed by gas condensation and fragmentation that ultimately lead to star formation, is a highly complex process. Nevertheless, we outline the key physical processes essential for star formation within collapsed haloes. Gas cooling leads to a drop in pressure, which, in turn, lowers the critical mass required for gravity and pressure to balance. This leads to fragmentation of the gas in the halo, where the gas clouds break up into even smaller and smaller clumps. However, this fragmentation process eventually stops when the gas becomes dense and optically thick, at which point the cooling timescale exceeds the free-fall timescale. The masses of the stars formed depend on when and how the fragmentation stops. The distribution of the stellar masses $m_*$ in a newly formed stellar population is usually characterized by the stellar initial mass function (IMF), defined as the relative number of stars born with mass in a given range. At very high redshifts, when the cooling was not very efficient in the absence of metals, the fragmentation stopped early and the stars formed tended to have much higher mass leading to a “top-heavy” IMF. These stars, formed from pristine gas, are expected to be extremely metal-poor and known as the \textit{Population III (Pop III)} stars. Due to their chemical composition, Pop III stars could only generate energy through nuclear fusion via the proton-proton (p-p) cycle, as they lacked the heavier elements required to fuel the more efficient CNO cycle. As a result, these stars were incredibly hot and possessed a very hard emission spectrum, producing far more energetic photons than the stars we see today, making Pop III stars highly efficient sources of ionizing photons in the early Universe.

Once these first stars begin to form, the energy released by them in the form of radiation or mechanical outflows can significantly impact the formation of subsequent stars through various ``feedback" effects. For instance, molecular cooling within minihaloes is extremely short-lived because hydrogen molecules are highly susceptible to dissociation by UV photons with energies between 11.26 and 13.6 eV, which are emitted by the very first stars. As molecular hydrogen is gradually depleted, star formation in the smallest minihaloes begins to be suppressed. Ultimately, this Lyman-Werner radiation background becomes so intense that Pop III star formation is entirely switched off in pristine minihaloes. Further episodes of star formation will then have to wait until the haloes grow large enough, attaining masses $\geq 10^8~\mathrm{M}_\odot$ so that the gas temperature crosses the threshold ($> 10^{4}$ K) for atomic cooling to become efficient. 

Additionally, given that Pop III stars are extremely massive, they have significantly shorter lifetimes (approximately 10$^6$ years), and end their lives in various types of supernova explosions depending on their initial mass. Among these, the most intriguing are the Pop III stars with masses in the range of 100 -  260 $\mathrm{M}_\odot$, which explode as Pair Instability Supernovae (PISN). These supernovae have exceptionally high metal yields, converting roughly half of their mass into heavy elements. Such energetic explosions, along with powerful stellar winds during the late stages of stellar evolution, enrich the surrounding gas with metals produced inside the cores of the first stars. As a consequence, species such as singly ionized carbon (CII) and neutral oxygen (OI) then help in increasing the cooling rate through excitation of their atomic fine-structure levels, allowing the gas to reach even lower temperatures and fragment into stars with significantly lower masses. This marks the onset of the more ``conventional" metal-enriched mode of star formation in haloes, leading to the formation of the second generation of stars, known as \textit{Population II (Pop II)} stars, whose IMF is dominated by low-mass stars and have a bottom-heavy shape. 

The amount of ionizing radiation a galaxy produces per unit time is primarily determined by the rate at which it forms stars and the properties of its stellar population (e.g., age, IMF, binarity, and metallicity). For example, the ionizing photon yield, for the same amount of stars, is higher at lower metallicities owing to the higher surface temperatures of such stars in the absence of metal-line cooling. Additionally, it is important to remember that feedback mechanisms arising from preceding stellar populations (whether Pop III or Pop II) would also impact the formation and evolution of the subsequent generation of stars, thereby affecting the emissivity of ionizing photons from a galaxy. Broadly speaking, these feedback processes are classified under the following three headings:

\begin{itemize}
    \item \textbf{Chemical feedback}: This refers to the change in the metallicity of the star-forming gas. The earliest metal-free stars synthesized heavy elements in their cores, which were eventually ejected into the surrounding medium through supernova explosions or stellar winds. The injection of metals alters the chemical composition of the gas, influencing the nature of new stars that are formed and consequently, the amount of ionizing photons they produce.
    
    \item \textbf{Mechanical feedback}: This refers to the injection of mechanical energy into the interstellar medium
     by massive stars in the form of stellar winds or supernova explosions. These energetic events can expel loosely bound gas from the potential well of haloes, either partially or completely, thereby suppressing star formation (and thus, the production of ionizing photons) — especially in low-mass systems.
     
    \item \textbf{Radiative feedback}: This refers to the effects caused by the heating of gas as a result of photoionization. Firstly, if the thermal velocity of the photoionized gas exceeds the escape velocity of the halo, the gas unbinds and can easily stream out from the potential well (a process known as photoevaporation), thereby depleting the gas reservoir in sufficiently low mass systems. Secondly, photoionization heating can also sufficiently raise the gas temperature to a point where its thermal pressure exceeds the gravitational binding energy, preventing gas from being accreted onto these systems.
\end{itemize}

Until now, we have only discussed the production of ionizing photons by stars within galaxies. However, not all of these photons produced will escape into the surrounding IGM. A portion of them may well be absorbed within the interstellar medium of the galaxy itself, most likely by dense clouds of neutral hydrogen gas. The escape fraction of ionizing photons, also known as Lyman continuum (LyC) photons, depends very sensitively on the density and clumping structure of neutral hydrogen and dust in the interstellar medium (ISM) and circumgalactic medium (CGM) of a galaxy. In the literature, two primary mechanisms of LyC leakage have been proposed - namely, an ionization-bounded nebulae with ``holes'', and density-bounded nebulae \citep{Zackrisson2013}. In the first scenario, it is believed that supernovae blast waves or stellar winds carve out low-density tunnels in the neutral ISM, allowing LyC photons to escape without being absorbed. This leads to an anisotropic leakage of ionizing radiation from the star-forming regions. In the second scenario, it is hypothesized that the LyC radiation from an intense starburst event depletes all of the surrounding neutral hydrogen before a full Strömgren sphere is developed, thereby allowing the ionizing radiation to escape more isotropically.

\subsection{Accreting black holes: Quasars}

Besides the stellar sources, a population of accreting black holes (namely, quasars) can also produce significant ionizing radiation and may have thus contributed to reionization. Observational studies indicate that the quasar luminosity function peaks at around $z \approx 3$, and exponentially declines at higher redshifts, raising considerable doubt about their contribution to reionization at higher redshifts ($z \geq 6$) \citep{Kulkarni2019}. Nevertheless, it turns out that several aspects of the reionization process change dramatically if quasars, rather than stars within galaxies, are the primary sources of ionizing radiation. 

Firstly, quasars have a much harder ionizing spectrum, producing a significantly higher number of high-energy photons compared to the softer UV radiation from stars. These highly energetic photons, often in the form of hard X-rays ($\geq$ 10 keV), have comparatively longer mean free paths and can thus travel much larger distances through the IGM. Therefore, they would ionize and heat the IGM much more uniformly than stellar radiation.  Moreover, these X-ray photons are capable of ionizing more than a single hydrogen atom: the first atom being ionized through regular photoionization, and several other atoms in its vicinity through repeated secondary collisional ionizations caused by the energetic photoelectrons generated from the primary photoionization. While secondary ionizations might seem to boost the ionization efficiency of quasars, this effect becomes less effective over time. As the ionized fraction in the IGM rises to a few percent, the number of secondary ionizations drops significantly and then, most of the energy from the primary photoelectrons goes into heating the medium through elastic Coulomb collisions. Therefore, lower-energy photons, from either quasars or stars, are still needed to achieve complete ionization of the Universe. 

The hard non-thermal spectra, which are characteristic of quasars, further suggest that they are capable of causing the double-reionization of helium (which requires photons with energies $>$ 54.4 eV, not produced in galaxies) almost simultaneously alongside the reionization of hydrogen. Additionally, some of these hard X-ray ($\geq$ 10 keV) photons, with mean free paths comparable to the horizon scale, are rarely absorbed and continue to free-stream redshifting to lower energy bands, and end up contributing to the soft X-ray background (XRB) which we observe at present. As a result, constraints on the unresolved soft (0.5–2 keV) X-ray background place strict limits on the allowed abundance of quasars and their contribution to reionization at $z \geq 6$.

The issue of the contribution of active galactic
nuclei (AGN) to reionization witnessed a renewed interest after claims of high number densities of faint AGN at $z \sim 4$ by \cite{Giallongo2019} and \cite{Boutsia2018}. When extrapolated to $z \approx$ 5.6, these findings suggest that AGN could have played a significant role in reionizing the Universe. Moreover, observational studies at low redshift ($z \approx 3–4$) show that the escape fraction of ionizing photons from quasars can be significantly higher compared to that from galaxies \citep{Grazian2018, Romano2019, Smith2020}. This picture of quasar-driven reionization has recently gained further observational support with the discovery of a population of moderate-luminosity AGN candidates, powered by early supermassive black holes, in deep JWST surveys at redshifts $4 \leq z \leq 13$ \citep{Harikane2023_AGN, Maiolino2023, Matthee2024}. Interestingly, the number densities of these JWST-detected AGN are comparatively higher ($\sim$ 10 times) than that expected from an extrapolation of the pre-JWST quasar luminosity function to these faint luminosities. In light of these latest observations, a recent analysis by \cite{Madau2024} suggests that AGN at $z>5$ could reionize the Universe without overproducing the unresolved X-ray background or contradicting HeII Lyman-alpha forest observations, provided their ionizing UV spectrum is exceptionally steep and/or their HeII-ionizing photons (energies $> 54.4~\text{eV}$) are internally absorbed.

\subsection{Exotic sources or mechanisms}

Besides the conventional astrophysical sources mentioned in the previous sub-sections, various other alternative mechanisms have also been proposed in the literature that could have contributed to the reionization of the IGM. One of the most widely discussed possibilities is the annihilation or decay of dark matter particles \citep{Mapelli2006, Belikov2009, Liu2016}. These dark matter interaction processes give rise to standard model particles, such as leptons and baryons, along with the emission of photons (whose exact energy depends on the particle masses involved) that can easily ionize the surrounding neutral atoms. Unlike conventional sources such as galaxies and quasars, ionizing photon production through such processes is not tied to the site of collapsed structures and can take place throughout the Universe, resulting in a more uniform reionization topology. Primordial black holes of intermediate mass (termed as ``mini quasars'') at $z > 10$ accreting gas through the Bondi-Hoyle process \citep{Madau2004, Ricotti2004}, or stellar-mass black holes in binary systems (``micro quasars'') accreting from their high-mass companion stars \citep{Mirabel2011}, have also been proposed as potential sources of ionizing photons. These early black holes may have contributed to the process of reionization before more conventional sources took over. There have also been studies that argue that a large fraction of stars in high-redshift galaxies probably formed in globular clusters \citep{Ricotti2002, Ma2021}, which leak their emitted radiation entirely into the IGM, thereby contributing a substantial fraction of photons for reionization.

\section{Observational probes of Reionization}\label{chap1:sec5}%

\subsection{Cosmic Microwave Background observations}

\begin{figure}[t]
\centering
\includegraphics[scale=0.5]{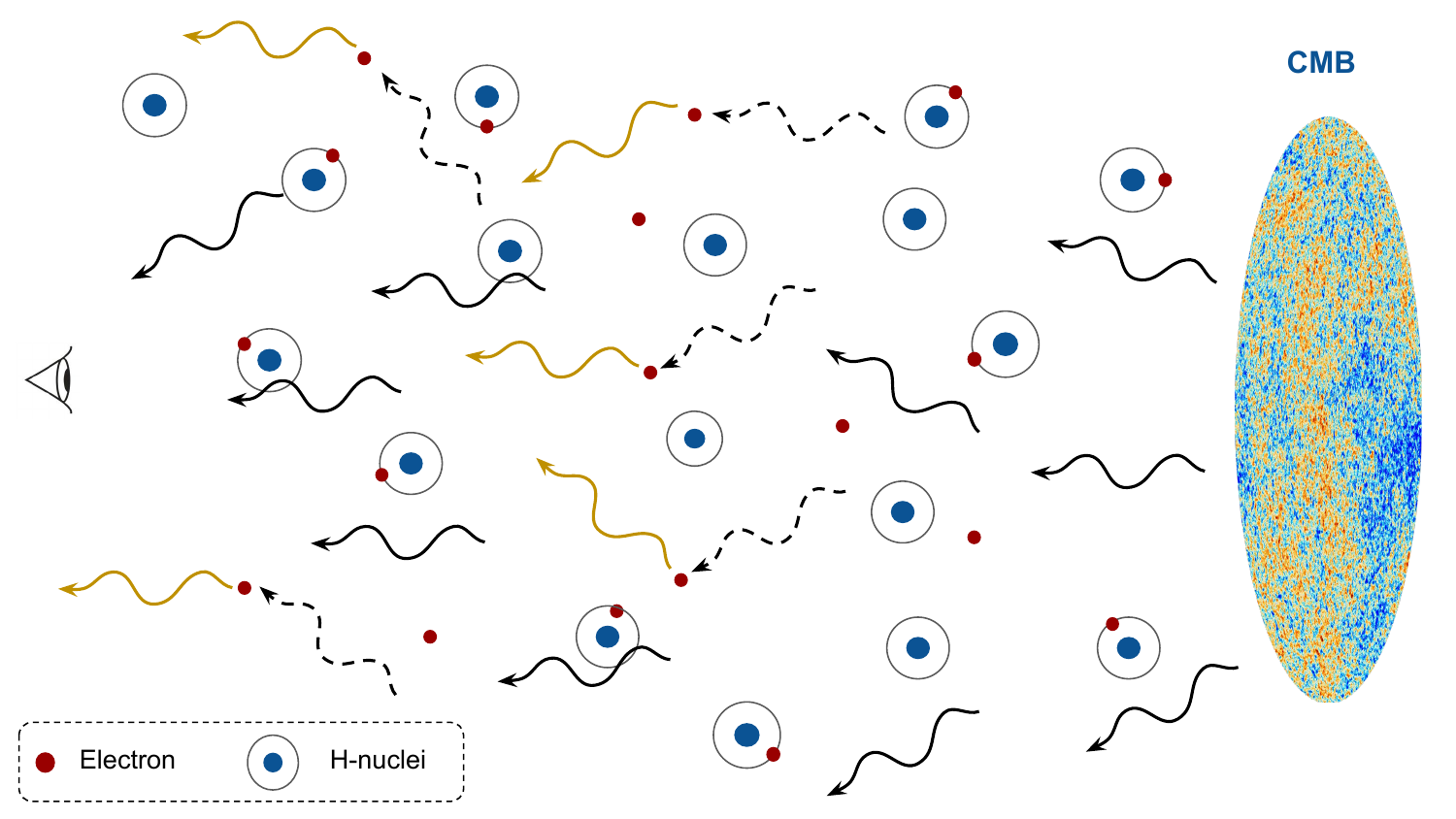}
\caption{A cartoon illustration showing the Thomson scattering of CMB photons by free electrons produced during reionization. While the majority of the photons (shown using black solid curves) from the epoch of recombination travel through the IGM unobstructed, some of the photons (shown using black dashed curves) are intercepted by free electrons on their way and get scattered away in random directions (shown using yellow curves). }
\label{fig:CMB}
\end{figure}

After the first atoms formed, primordial radiation decoupled from matter and traveled mostly unimpeded through the Universe. Today, we observed these photons, originating from the epoch of the last scattering, as the cosmic microwave background (CMB). However, once the first luminous sources began ionizing the IGM, these CMB photons were re-scattered on their way through the IGM by the free electrons produced during reionization, see \fig{fig:CMB} for an illustration. This scattering reduces the amplitude of the primary temperature anisotropies on angular scales smaller than the horizon scale at the re-scattering epoch by a factor of  $e^{-\tau_{\rm el}}$, where
\begin{equation}
    \tau_\mathrm{el} = c~\sigma_T~\bar{n}_H \int_{t_0}^{t_\mathrm{CMB}} dt~\chi_\mathrm{He}(t)~Q_\mathrm{HII}(t)~a^{-3}(t)
\end{equation}
is the integrated electron scattering optical depth of CMB photons. In the above equation, $\sigma_T$ is the Thomson scattering cross-section, and the integral runs from the present epoch $t_0$ to the origin of the CMB $t_\mathrm{CMB}$. The combination $\bar{n}_H~\chi_\mathrm{He}~Q_\mathrm{HII}~a^{-3}$ represents the density of free electrons, and ensures that its contribution to the integral is essentially zero at epochs before the onset of reionization. As a result of this scattering, the angular power spectrum of CMB temperature anisotropies is damped by a factor of $e^{-2\tau_{\rm el}}$ on these small angular scales (i.e., high-$\ell$), while remaining largely unaffected on larger scales. Unfortunately, this damping is completely degenerate with the amplitude ($A_s$) of the primordial power spectrum of scalar perturbations and is extremely difficult to discern directly from observations.

Thankfully, the Thompson scattering process also causes the scattered radiation to become linearly polarized when free electrons interact with the incident quadrupolar CMB radiation field. This secondary curl-free (E-mode) polarization is most pronounced on angular scales corresponding to the horizon size at the epoch when the photons were scattered. Therefore, the detection of polarization on scales larger than the angle subtended by the causal horizon at the surface of last scattering (i.e., during cosmological recombination) provides strong evidence for Thomson scattering by free electrons in the post-recombination Universe. This feature in the polarization power spectrum at low angular multipoles (low-$\ell$) serves as a tell-a-tell sign of cosmic reionization and is generally called the 'reionization bump'. Additionally, similar to the temperature anisotropies, the polarization anisotropies are also suppressed by a factor of $e^{-\tau_{\rm el}}$ at angular scales larger than the horizon size at the epoch of reionization. Moreover, the location of the reionization bump and its amplitude in the polarization power spectrum depends on when the Universe was reionized and the square of the electron scattering optical depth ($\tau_{\rm el}$). Hence, the shape of the large-scale CMB polarization power spectrum helps in breaking the degeneracy between $A_s$ and $\tau_{\rm el}$, which is expected to exist in measurements of temperature anisotropies as described above. Calculating the value of $\tau_{\rm el}$ from observations of the CMB temperature (T) and polarization (E) auto-correlation and cross-correlation power spectra requires adopting a model for the redshift evolution of the number density of free electrons, or equivalently, a model for reionization. For instance, the latest measurement of the electron scattering optical depth at $\tau_{\rm el}=0.054 \pm 0.007$ reported by \citet{Planck2018} has been obtained from a combined analysis of the full Planck mission TT ($2 \leq \ell \leq 2500$), TE ($30 \leq \ell \leq 2000$) and EE ($2 \leq \ell \leq 2000$) data and the Planck CMB lensing data ($8 \leq \ell \leq 400$), assuming a tanh model for reionization, which is parametrized by a reionization midpoint $z_{\mathrm{re}}$ and a redshift width ${\Delta}z_{\rm re}$ (kept fixed to a value of ${\Delta}z_{\rm re} = 0.5$)\footnote{In the tanh parameterization of reionization, the evolution of the globally averaged free electron fraction $x_e$ as function of redshift ($z$) is given by the following mathematical expression -  $x_e(z)= \chi_\mathrm{He}(z)~Q_\mathrm{HII}(z) = \frac{1}{2}\left[1+\tanh\left(\dfrac{y(z_{\rm re})-y}{\Delta_y}\right)\right]$, where $y(z)=(1+z)^{3/2}$, $\Delta_y = 1.5 \, \sqrt{1+z_{\rm re}} \, {\Delta}z_{\rm re} $ and other symbols have their usual meaning}.  In this case, the measured value of $\tau_{\rm el}$ corresponds to a reionization ``midpoint'' of $z_{\mathrm{re}} = 7.7 \pm 0.7$. However, it is important to realize that this simplified redshift-symmetric tanh model of reionization, widely adopted in the standard analysis of CMB data, greatly differs in shape from both physical and empirical models of reionization history, which are based on hierarchical models of structure formation and/or inferred from other high-redshift astrophysical observables \citep{Paoletti2024}. Moreover, the value of the optical depth, in principle, could also depend on the details of the reionization history assumed (e.g., see \citet{Hazra2017, Miranda2017, Qin2020, Planck2018, Chatterjee2021_CosmoReionMC}).

Reionization also produces additional temperature anisotropies in the CMB spectrum on small scales which offer complementary information about the process compared to large-angle polarization measurements that are largely insensitive to its patchiness and duration. The bulk motion of free electrons in ionized bubbles relative to the CMB rest frame causes a Doppler shift in the scattered photons, leading to a small shift in the observed CMB temperature. This phenomenon is known as the kinetic Sunyaev-Zel’dovich (kSZ) effect.  In the non-relativistic limit, the temperature change is proportional to the line-of-sight bulk velocity of the electrons and the number density of free electrons. We emphasize that the kSZ effect (which arises from the bulk motion of free electrons) is fundamentally different from the thermal Sunyaev-Zel'dovich (tSZ) effect, which arises due to the Compton scattering of CMB photons inside hot gas residing in large-scale structures such as galaxy clusters. 

The kSZ signal broadly encompasses two distinct components, the homogeneous kSZ and the patchy kSZ. The homogeneous kSZ signal arises from spatial inhomogeneities in the baryonic density in the fully ionized universe. In contrast, the patchy kSZ signal is driven by fluctuations in the ionization fraction during the EoR. The amplitude of the patchy kSZ power spectrum is primarily determined by the timing and duration of reionization, while its shape depends on the size distribution of the ionized bubbles. Several studies in the literature have discussed the prospects of using the patchy kSZ signal for inferring not only the mid-point and duration of reionization but also the sources driving the process \citep{Park2013, Gorce2020, Paul2021, Nikolic2023, Jain2024}. Recently, the first $3\sigma$ detection of the kSZ signal was reported by \cite{Reichardt2021} using the temperature and polarization data (measured across the multipole range - $2000 \leq \ell  \leq 11,000$) from the SPT-SZ and SPT-pol surveys, yielding a value of $D^{\mathrm{kSZ}}_{\ell}=\ell(\ell+1)C^{\mathrm{kSZ}}_{\ell} =3 \pm 1 ~\mu \mathrm{K}^2$ at an angular multipole of $\ell=3000$. 

\subsection{Lyman-alpha absorption studies}

\begin{figure}[t]
\centering
\includegraphics[scale=0.5]{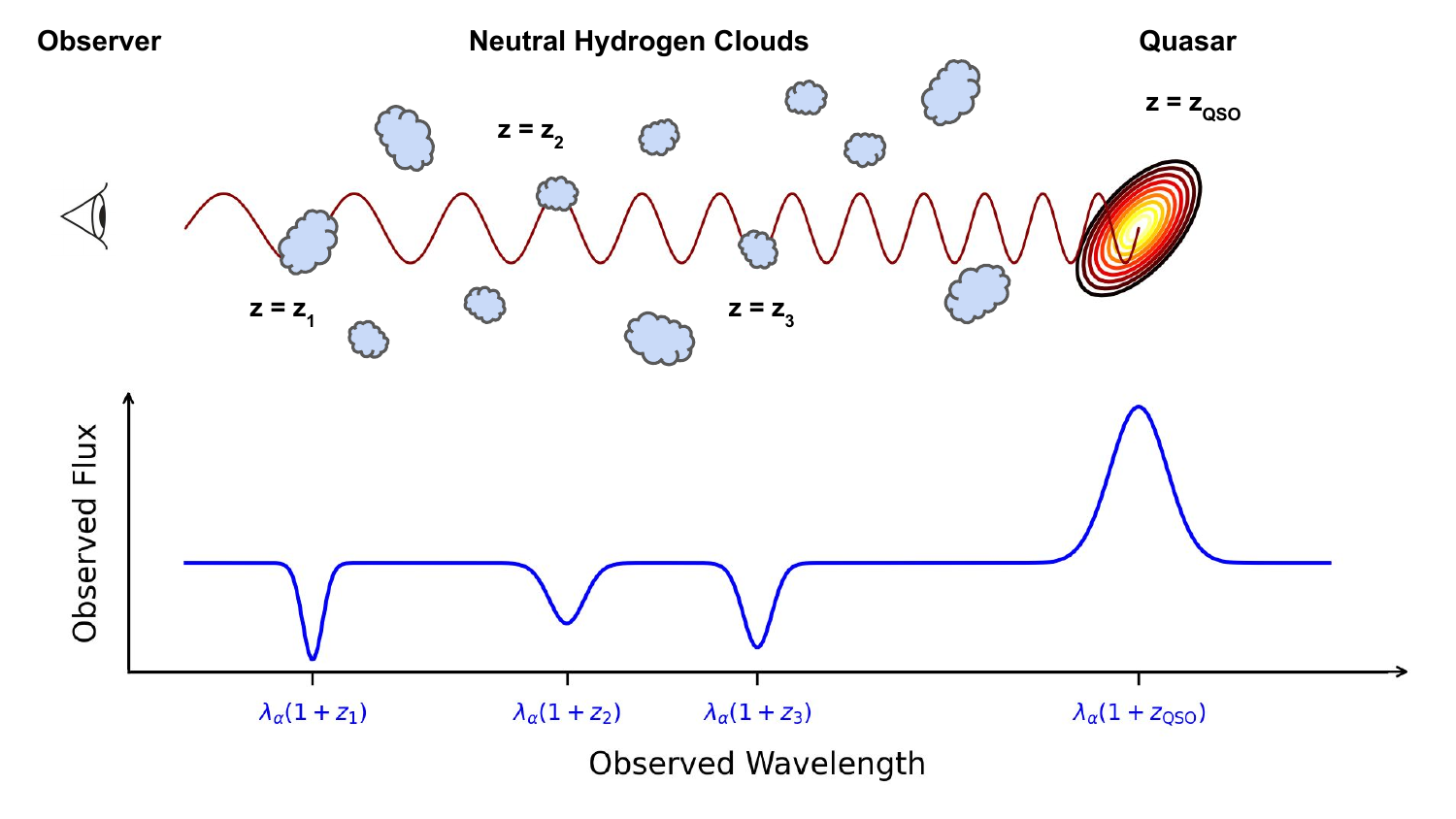}
\caption{A cartoon illustration of the Lyman-$\alpha$ forest observed in the spectra of distant sources. This arises due to the absorption of the radiation emitted by a background source as it passes through multiple clouds of neutral hydrogen along the line of sight. }
\label{fig:lya_forest}
\end{figure}

Absorption lines seen in the spectra of distant astronomical sources provide an elegant way to study the physical state of the intervening medium at different redshifts along a particular line of sight. Among the various sources known to exist in the early Universe, quasars are believed to emit strongly across a wide range of wavelengths, with their intrinsic spectrum showing broad emission lines superimposed on a smooth continuum. As these emitted photons travel towards us, their wavelengths are stretched due to the expansion of the Universe. Eventually, photons that were emitted with rest-frame wavelengths shorter than the Lyman-$\alpha$ wavelength ($\lambda_\alpha$ = 1216 $\angstrom$) will get redshifted close to the Lyman-$\alpha$ transition wavelength at some point on their way. If this occurs at a region in the IGM where there are substantial neutral hydrogen atoms, the atoms will absorb these photons, transitioning from the ground state to the first excited state (i.e., undergoing the 1s $\rightarrow$ 2p Lyman-$\alpha$ transition) and then again de-excite, re-emitting the photons in random directions. This resonant scattering results in an absorption line in the observed quasar spectra blueward of the rest-frame Lyman-$\alpha$ emission. For an absorption system at some intermediate redshift $z_\mathrm{abs}$ (where, 0 $<$ $z_\mathrm{abs} < z_{\mathrm{source}}$) along the line of sight, we would expect absorption signatures in the observed spectrum at a wavelength of (1 + $z_\mathrm{abs}$)$\lambda_\alpha$. The presence of multiple intergalactic clouds of neutral hydrogen lying at different redshifts along the path would then leave a series of narrowly spaced absorption lines in the observed spectrum, which are collectively referred to as the Lyman-$\alpha$ forest (see \fig{fig:lya_forest} for an illustration). On the other hand, photons emitted with rest-frame wavelengths $\lambda > \lambda_\alpha$ will have their wavelengths redshifted further and further away from the Lyman-$\alpha$ resonance as they travel toward us, and therefore will never be absorbed by any intervening neutral hydrogen. As a result, a sharp decrease in flux is expected across the position of Lyman-$\alpha$ emission line in the observed spectrum if there is neutral hydrogen present between the quasar and us. This is known as the \emph{Gunn-Peterson (GP) effect} \citep{Gunn1965}. It should be emphasized that imprints of Lyman-$\alpha$ absorption features due to a neutral IGM on observed spectra are not limited to quasars; this effect applies equally to any distant background source, such as primeval galaxies and gamma-ray burst afterglows.

The key observable in quasar absorption spectra is the transmitted flux $F_\alpha = e^{-\tau_\alpha}$, where
\begin{align}
\label{eq:tau_alpha}
\tau_\alpha (\lambda_{\rm obs},z) &= \frac{\sigma_{\alpha}~c}{H(z)}~\xHI~n_H~(1 + z)^3
\nonumber \\
& \simeq 2.3 \times 10^{5} \, \xHI \left(\frac{n_H}{\bar{n}_H}\right) \left(\frac{\Omega_{\rm b}h^{2}}{0.022}\right)\left(\frac{\Omega_{\rm m}h^{2}}{0.142}\right)^{-1/2} \left(\frac{1-Y}{0.76}\right)\left(\frac{1+z}{5}\right)^{3/2}
\end{align}
is the optical depth, due to Lyman-\(\alpha\) absorption at the absorption redshift $z$ ($ \equiv \lambda_{\rm obs}/ \lambda_{\alpha} -1)$, observed today at a wavelength of $\lambda_{\rm obs}$. In the above expression, $\sigma_\alpha$ is the Lyman-$\alpha$ absorption cross-section, $\xHI$ is the neutral hydrogen fraction, and $H(z)$ is the Hubble parameter. In deriving these expressions, we have neglected the width of the absorption cross-section and effects due to peculiar velocities for simplicity. We have also approximated $H(z)\simeq H_{0}\Omega_{\rm m}^{1/2}(1+z)^{3/2}$ in the second expression, valid for $z \gtrsim 2$.  When the IGM is highly ionized, the residual neutral fraction $\xHI$ can be related to the thermal state of the IGM and the background photoionization rate $\Gamma_\mathrm{HI}$ assuming an equilibrium between photoionization and recombination, valid for low-density gas that forms the Lyman-$\alpha$ forest - 
\begin{equation}
\label{eq:xHI_from_PIE}
x_{\rm HI}  = \frac{\alpha_R(T)~ \chi_\mathrm{He} ~ n_H~(1 + z)^3}{\Gamma_{\mathrm{HI}}}.
\end{equation}

While the global neutral hydrogen fraction $\bar{x}_\mathrm{HI}$ can be directly inferred from the average optical depth $\bar{\tau}_\alpha$, we remind the readers that the main observable is the average transmitted flux $\bar{F}_\alpha \neq e^{-\bar{\tau}_\alpha}$, complicating direct inferences of $\bar{x}_\mathrm{HI}$ from quasar observations. Nonetheless, it is clear from \eqn{eq:tau_alpha} that Lyman-$\alpha$ absorption is highly sensitive to the neutral hydrogen density. For instance, a neutral hydrogen fraction of $x_\mathrm{HI} \approx 10^{-4}$ results in $\tau_\alpha \gg 1$, causing the transmitted flux to approach zero ($F_\alpha \to 0$). This results in extended absorption ``troughs'' blueward of the Lyman-$\alpha$ emission in the observed spectrum. The absence of a GP trough only provides an upper limit on $x_{\mathrm{HI}}$. Observations showing no such troughs at $z \lesssim 5$ suggest the IGM is highly ionized at these redshifts. Thus, the Lyman-$\alpha$ forest is a powerful probe of the late stages of reionization, when the IGM is predominantly ionized with only minor amounts of residual neutral hydrogen. However, for higher neutral fractions $\xHI \gg 10^{-3}$, the quasar absorption spectra show minimal transmitted flux, making interpretation of the observations at higher redshifts challenging without detailed theoretical models. 

Recent spectroscopic observations of high-redshift quasars at $5 \lesssim z \lesssim 6$ have revealed significant variation in the average transmitted flux across different sight lines \citep{Becker2015, Bosman2018, Bosman2022}. These fluctuations are larger than those expected in standard models with uniform ionizing backgrounds and power-law temperature-density relations. As a result, a number of models that invoked fluctuations in the temperature \citep{DAloisio2015}, background photoionization rate \citep{Davies2016} and density distribution \citep{Chardin2017} were proposed. However, a delayed end to reionization at $z  \approx$  5 seemed to most satisfactorily explain the observed scatter in the transmitted flux, as such a scenario would leave large ``islands'' of neutral hydrogen even at $z <$ 6 that could cause significant absorption along certain quasar sight lines \citep{Kulkarni2019_IGM, Nasir2020}. 

A complementary method of constraining the neutral fraction towards the end of reionization is by analyzing the ``dark pixels'', which entails measuring the fraction of pixels that are entirely absorbed in the observed Lyman-$\alpha$ forest spectra. This offers a relatively model-independent estimate of the filling factor of ionized regions \citep{McGreer2015, Jin2023}. However, this method gives only an upper limit on $\bar{x}_{\rm HI}$, as residual HI inside ionized regions (in place of neutral regions in the IGM) may also saturate the Lyman-$\alpha$ forest.

Since the reionization process is patchy, small highly-ionized regions are expected to exist even when the Universe is still largely neutral. These regions would manifest as sharp ``spikes'' of transmission in the spectra of high-redshift quasars. The position and spacing of these spikes, interspersed with regions of no detectable flux (known as ``dark gaps''), are expected to hold more information than what can be captured by just measuring the effective optical depth. Studies have shown that the number and height of these transmission spikes are sensitive to the ionization fraction of the IGM, which in turn depends on the photoionization rate and the temperature of the gas in the ionized regions \citep{Gaikwad2020}. On the other hand, the size distribution of these so-called dark gaps has also been claimed to be a robust diagnostic of the ionization state of the IGM during late stages of reionization \citep{Gallerani2006, Gallerani2008, Zhu2021}.  However, these methods of estimating $\xHIbar$ are usually model-dependent, requiring several assumptions to be made. 

Quasars not only act as backlights for absorption studies but also create extensive ionized regions around themselves, where the strength of the ionizing radiation field is higher than average. These regions, known as proximity zones, are the only areas where significant transmission occurs just blueward of the quasar's  Lyman-$\alpha$ emission. If the sizes of these proximity zones can be reliably measured from absorption spectra, they could provide information about the amount of neutral hydrogen in the IGM around the quasar \citep{Fan2006, Bolton2007, Carilli2010}. However, inferences of $\xHI$ drawn from observed quasar proximity zones depend on associated assumptions regarding the intrinsic properties of quasars, such as their ionizing luminosity and age.

Observations of the mean Lyman-$\alpha$ transmitted flux, combined with numerical simulations, have also helped estimate the HI photoionization rate $\Gamma_{\mathrm{HI}}$. In such studies, one essentially tries to match the mean transmission in the simulated spectra to the observational measurements at $5 < z < 6$, since we expect $\tau_\alpha \propto \Gamma_{\mathrm{HI}}^{-1}$ to hold in photoionization equilibrium (see \eqns{eq:tau_alpha}{eq:xHI_from_PIE}). However, the photoionization rate estimated through such a procedure depends on the assumptions made regarding the mean free path $\lambda_\mathrm{mfp}$ of ionizing photons and the temperature-density relation in the low-density IGM. In recent times, it has also been possible to measure $\lambda_\mathrm{mfp}$ directly from quasar absorption spectra, which suggests that reionization is likely to complete only by $z \sim 5.3$ \citep{Zhu2023}. Such measurements of $\lambda_\mathrm{mfp}$ and $\Gamma_{\mathrm{HI}}$ can be used to obtain an estimate of the number of ionizing photons available in the IGM (since, $\dot{n}_\mathrm{ion} \propto  \Gamma_{\mathrm{HI}}^{-1} ~  \lambda_{\rm mfp}$) \citep[e.g.,][]{Gaikwad2023}.

Reionization can also be constrained through the thermal history of the IGM, measured using quasar absorption spectra. The power spectrum of Lyman-$\alpha$ absorption fluctuations shows small-scale suppression due to pressure smoothing, which depends on the thermal evolution of the gas \citep{Rorai2017, Walther2019}. Recent studies \citep[e.g.,][]{Gaikwad2020} have measured the IGM temperature using transmission spikes at $z \sim 5.5$, allowing constraints on the reionization history through comparisons with a theoretical model \citep{Maity2022}.

\subsection{High-redshift galaxies: Lyman-break galaxies and Lyman-alpha emitters}

As mentioned previously, the high volume densities of star-forming galaxies at high redshifts make them the most favored source of ionizing photons. Therefore, to reliably trace the course of reionization, it is essential to first obtain a census of the Lyman continuum photons contributed by them for ionizing hydrogen in the IGM. However, directly measuring the ionizing radiation produced from such early galaxies escaping into the IGM is extremely challenging if not impossible. This difficulty arises due to the presence of substantial amounts of neutral hydrogen gas in their ISM and the IGM, both of which absorb the produced LyC radiation. 

Because of these difficulties, the total ionizing radiation contributed by galaxies for reionization is usually estimated indirectly by combining three different pieces of information, namely, the rest-frame UV luminosity density, an efficiency factor $\xi_{\rm ion}$ that translates the non-ionizing UV luminosity of galaxies to their Lyman-continuum radiation, and the escape fraction $\fesc$.\footnote{Strictly speaking, this escape fraction $\fesc$ refers to the relative fraction of ionizing photons escaping from galaxies to the fraction of 1500 $\angstrom$ UV-continuum photons that escaped.} Without any loss of generality, we can write the ionizing emissivity of galaxies which appears as the source term in \eqn{eqn:Madau_Eqn} to be,

\begin{align}
    \dot{n}_\mathrm{ion} \equiv \frac{d n_{\rm ion}}{d t} &= \int d\LUV~\Phi (\LUV)~\dot N_{\rm ion}(\LUV) ~\fesc(\LUV), \nonumber \\
     &= \int d\LUV~\Phi (\LUV)~\LUV~\xi_{\rm ion}(\LUV) ~\fesc(\LUV),
\label{eq:niondot}
\end{align}
where $\dot N_{\rm ion} \equiv \LUV \, \xi_{\rm ion}$ represents the number of ionizing photons produced per unit time by a galaxy with UV luminosity $\LUV$ and $\Phi (\LUV)$ denotes the UV luminosity function of galaxies, which quantifies the abundances of galaxies in a given comoving volume per unit luminosity interval. It must be noted that the integral above formally extends down to the faintest of galaxies that are capable of producing ionizing photons. These ultra-faint galaxies are unlikely to be individually detectable in current or future galaxy surveys and therefore the observed UV luminosity function would need to be extrapolated down to lower luminosities, introducing uncertainties to the contribution of faint galaxies towards reionization.    

Over the past decade, the commissioning of large galaxy surveys using space-based and ground-based telescopes like the James Webb Space Telescope (JWST), the Hubble Space Telescope (HST), the Spitzer Space Telescope, the Subaru Telescope \citep{Bouwens2015, Bouwens2017, Atek2018, Ono2018, Bowler2020, harikane2022, Donnan2023, Harikane2023, Bouwens2023, McLeod2024, Donnan2024, Whitler2025} has resulted in a growing sample of high redshift galaxies. A widely used method for detecting high-redshift galaxies leverages the distinctive spectral break caused by the absorption of emitted photons by intergalactic and interstellar neutral hydrogen along the line of sight. Therefore, the light emitted blueward of a Lyman-series transition has an extremely low probability of reaching us, making the galaxy distinct in different broadband filters of the telescope. Galaxies selected using this technique are called the Lyman-break galaxies (LBGs). These galaxy observations have been instrumental in obtaining robust measurements of the rest-frame UV luminosity function $\Phi (\LUV)$ of galaxies at high redshifts ($6 < z < 10 $) and more recently, even out to redshifts as high as $z \approx 15$.

The second most crucial ingredient for calculating the ionizing photon budget from galaxies is the production efficiency of ionizing radiation ($\xi_{\mathrm{ion}}$), which depends on the spectral energy distribution of the stellar population in high-$z$ galaxies. Several theoretical stellar population synthesis (SPS) studies \citep{Liu2024, Wilkins2016}  have shown that the number of ionizing photons produced depends strongly on the choice of the SPS model (isolated or binaries) and astrophysical factors such as stellar age, stellar metallicity, stellar IMF, mode of star-formation etc. On the observational front, with the growing number of line-emitting galaxies and advancements in spectroscopy, it has been possible to obtain a lower bound on $\xi_{\mathrm{ion}}$ based on the measured intensity of nebular emission lines (such as Balmer lines) in galaxy spectra along with certain assumptions about the recombination process. Using JWST observations of high-$z$ galaxies, several studies initially inferred values of $\log_{10} \big[\xi_{\rm ion}/({\rm ergs}^{-1}\ {\rm Hz}) \big]$ in the range of $25.5-26.0$ \citep{CurtisLake2023, Rinaldi2023, Endsley2023, Atek2024_Spectroscopy, Simmonds2024, AlvarezMarquez2024, Calabro2024}, which were considerably higher than the canonical values $\log_{10} \big[\xi_{\rm ion}/({\rm ergs}^{-1}\ {\rm Hz}) \big] \approx 25.2$ obtained from previous studies with the HST \citep{Robertson2015} (though, also see \citet{Matthee2023, Simmonds2024_p2, Begley2025}). Some of these studies have also reported an increasing trend of $\xi_{\mathrm{ion}}$ towards higher redshifts and fainter galaxies. 

The last but perhaps the most important ingredient for reionization history calculations is the escape fraction, which is completely degenerate with $\xi_\mathrm{ion}$. A direct observational estimate of $\fesc$ during reionization is impossible for the reasons outlined earlier in this section. Consequently, indirect estimators of $\fesc$ are used - mostly, derived from detailed studies of correlations between LyC leakage and various galaxy properties in low-redshift analogs of LyC leakers \citep{Flury2022, Choustikov2024}. Some of the different diagnostics proposed include high $[\mathrm{OIII}]/[\mathrm{OII}]$ (O$_{32}$) ratios \citep{Jaskot2013, Nakajima2014, Izotov2018}, UV continuum slope \citep{Chisholm2022}, velocity separation between Lyman-$\alpha$ peaks \citep{Verhamme2015, Verhamme2017, Izotov2018, Izotov2021}, Lyman-$\alpha$ equivalent width \citep{Steidel2018,Pahl2021,Begley2022} and so on. However, it must be remembered that many of these indirect estimators could suffer from degeneracies related to stellar properties like age, metallicity, etc.

Combining these, it becomes straightforward to obtain the total ionizing emissivity of galaxies at a given redshift and also study the relative contribution of galaxies towards this photon budget as function of luminosity.  These observations therefore hold the promise of resolving the long-standing debate about the contribution of low-luminosity galaxies to the ionizing photon budget for reionization. In the literature, some studies \citep{Sharma2016, Naidu2020, Joshi2024} contend that faint galaxies (having magnitude $M_\mathrm{UV} \geq$ -17) make only a negligible contribution, while others \citep{Anderson2017, Lewis2020, Atek2024_Spectroscopy, Chakraborty2024, Simmonds2024} argue that these sources play a dominant role in the reionization of the Universe. 

Given the intricate interplay between cosmic reionization and galaxy formation, it is natural to expect that cosmic reionization would also leave noticeable imprints on both the physical and statistical properties of galaxies. In addition to shedding light on the contribution of different galaxies to the reionization photon budget via \eqn{eq:niondot}, these galaxy surveys offer crucial insights into star formation activity and the properties of stellar populations in early galaxies. For instance, the rest-frame UV continuum emission (1250–1500 $\angstrom$), which gets redshifted into optical wavelengths by the time it reaches us, serves as a measure of its instantaneous star formation rate, since UV radiation is primarily emitted by massive, hot stars that are short-lived. Another higher-order summary statistic that has recently become available at high redshifts from such surveys is the two-point correlation function which quantifies the clustering of a galaxy population in two- or three-dimensional space. The topology and morphology of reionization depend crucially on how the ionizing sources are distributed in space. Star formation and ionizing photon production are also influenced by feedback mechanisms, including radiative feedback from cosmic reionization. This feedback, which heats gas and suppresses star formation in low-mass haloes, is expected to flatten or turn over the faint end of the galaxy UV luminosity function. Any detection of such a feature in the luminosity function will help better understand the feedback mechanisms.

A complementary approach which is now increasingly used for detecting star-forming galaxies at high redshifts is based on the identification of redshifted line emission within a narrow redshift range from narrow-band imaging or spectroscopic observations. Galaxies that are selected using this technique are called line-emitting galaxies. Among the various line emissions expected from high-redshift galaxies, the Lyman-$\alpha$ line, which results from the recombination of ionized hydrogen in star-forming regions, is perhaps the most promising for such searches owing to its high intrinsic strength. Galaxies selected using their strong Lyman-$\alpha$ emission are known as Lyman-alpha emitters (LAEs).

The sensitive dependence of Lyman-$\alpha$ transmission on the presence of neutral hydrogen ($\mathrm{HI}$) around galaxies or in the IGM makes LAEs a promising probe of cosmic reionization \citep{Dijkstra2014, Ouchi2020}.  Galaxies are believed to form in overdense regions of the Universe, which are also the first to be reionized. Consequently, Lyman-$\alpha$  photons emitted by galaxies within HII regions can traverse through the ionized bubble, redshifting away from line resonance, before encountering the neutral IGM. Due to the strong frequency dependence of the Lyman-$\alpha$ absorption cross-section, these emergent photons are then less likely to be scattered out of the line of sight, allowing a substantial fraction of them to reach the observer. Therefore, crudely speaking, the visibility of galaxies in Lyman-$\alpha$ can be used to draw inferences about the presence of neutral hydrogen in the intervening medium.

With the progress of reionization, larger HII regions form around galaxies, increasing the prospects of detecting a higher number of galaxies emitting in Lyman-$\alpha$. As a result, the Lyman-$\alpha$ luminosity function of LAEs is expected to show a sharp drop in amplitude with increasing redshift (where the IGM becomes increasingly neutral). This distinctive evolution in the abundance of LAEs, as opposed to the LBGs over similar redshifts, can help in constraining the timing of reionization, assuming that the intrinsic Lyman-$\alpha$ properties of galaxies do not significantly change. Given the fact that the Lyman-$\alpha$ emission from galaxies within large ionized bubbles remains (relatively) unattenuated compared to that from galaxies located elsewhere (such as inside small ionized bubbles or neutral regions), it is only to be expected that the narrow-band Lyman-$\alpha$ searches will preferentially detect galaxies residing in the ionized regions of the IGM. Therefore, the ``observed'' spatial distribution of LAEs is expected to be modulated by the large-scale ionization field. With the ionized regions becoming increasingly rare at higher redshifts, one anticipates that the clustering of LAEs (quantified by the amplitude of its two-point correlation function) should also increase with increasing redshift. Unfortunately, the interpretation of observations of LAEs at high redshifts is not straightforward and plagued by several uncertainties such as the amount of relative velocity shift of the Lyman-$\alpha$ emission line as it emerges from the source, the density distribution and the size of ionized regions around these sources, etc., thereby requiring detailed theoretical models. 

In the last decade, sensitive narrowband surveys with the Subaru Telescope have been successful in increasing the sample of high-redshift LAEs, enabling robust determination of the Lyman-$\alpha$ LFs and LAE clustering, which agree well with the theoretical expectations outlined above \citep{Konno2018, Itoh2018, Goto2021}. More recently, observations with the JWST have pushed the boundaries of high-$z$ LAE studies even further by facilitating the detection of numerous LAEs and allowing for their detailed characterization deep into the EoR \citep{Witstok2024, Tang2024, Jones2024}. The most exciting result of the lot has been the detection of Lyman-$\alpha$ emission from galaxies at redshifts as high as $z$ = 10.6 \citep[GNz-11;][]{Bunker_z10p6} and $z$ = 13 \citep[JADES-GS-z13-1-LA;][]{Witstok2024_z13}, when the Universe is significantly neutral. Some of the latest studies \citep[e.g.,][]{Endsley2022, Whitler2024} have suggested the presence of large galaxy overdensities around some of the stronger Lyman-$\alpha$ emitters at $z > 6$, lending support to the notion that such overdensities would be easily able to source sufficiently large-sized ionized bubbles, which in turn would facilitate Lyman-$\alpha$ transmission.

\subsection{21~cm Emission}

The Lyman-$\alpha$ transition presents a significant hurdle in probing the early and intermediate stages of reionization since even a tiny amount of neutral hydrogen (around 1 part in 100,000) can make the IGM entirely opaque to Lyman-$\alpha$ radiation. Similarly, CMB probes are integrated measurements of the ionization state along the line of sight, offering no insights into the intricate details of the reionization process. In this context,  an attractive alternative to probe the different stages of reionization is the so-called 21~cm radiation arising from the hyperfine transition in the ground state of neutral hydrogen atoms. More specifically, when the spins of the proton and electron in a neutral hydrogen atom switch between parallel (triplet state) and anti-parallel (singlet state) configurations, the atom emits or absorbs a photon with a rest-frame wavelength of 21.1 cm (i.e., equivalent to a rest-frame frequency of 1420.4 MHz). However, the strength of the 21~cm line transition is extremely weak because the probability of spontaneous emission is low, making them harder to detect unless they are produced from regions with a high number of neutral atoms or those that are exceptionally dense. The optical depth ($\tau_{21}$) of the diffuse IGM towards 21~cm radiation at a redshift $z$ is thankfully extremely low - several orders of magnitude smaller than that of Lyman -$\alpha$ photons.

\begin{figure}[t]
\centering
\includegraphics[scale=0.5]{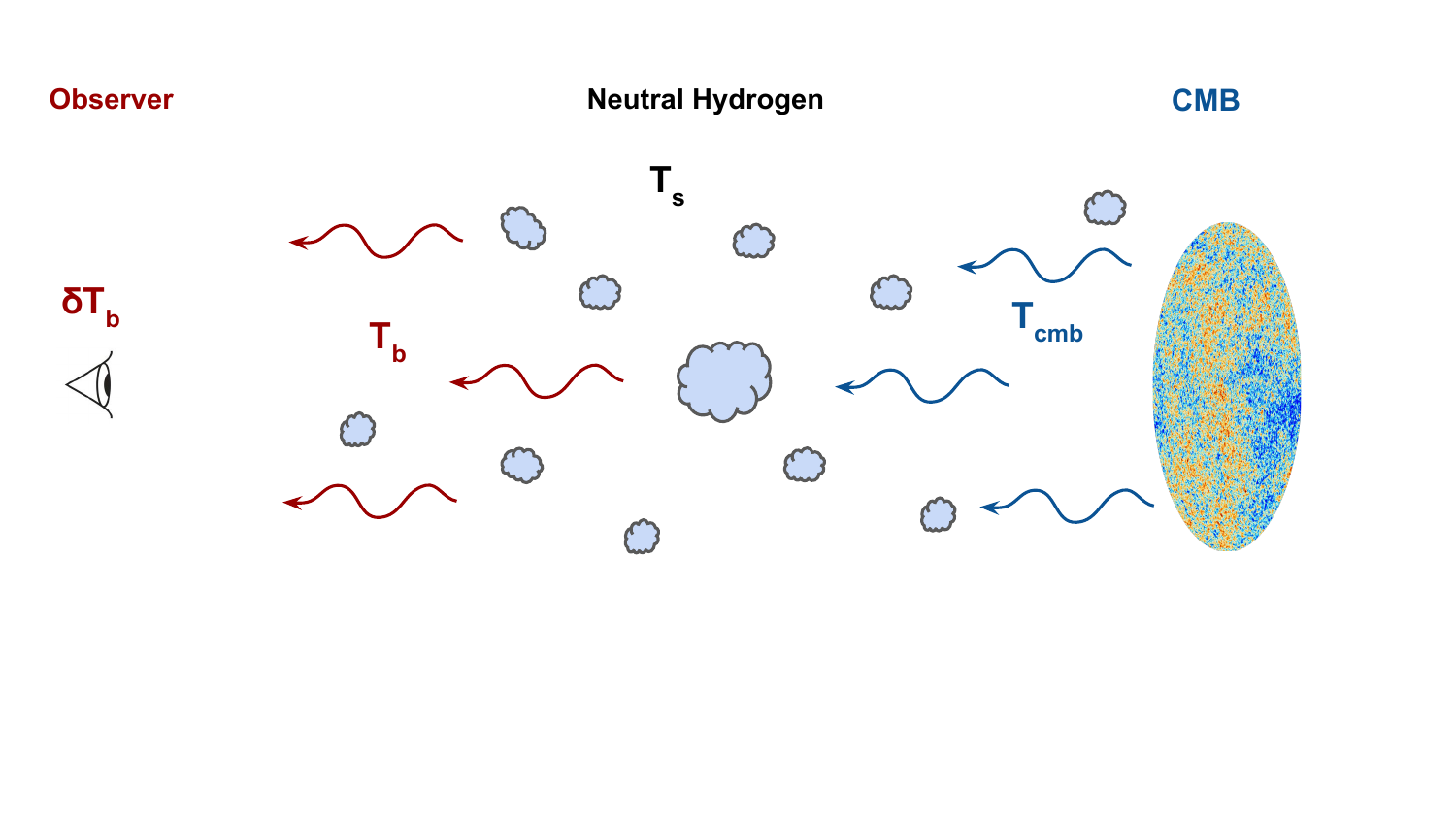}
\caption{A cartoon illustration showing the emission or absorption of 21~cm photons by intergalactic patches of neutral hydrogen with a spin temperature of $T_S$, when irradiated by a background radiation (CMB) having brightness temperature of $T_{\rm CMB}$.}
\label{fig:21cm}
\end{figure}

When background radiation passes through a patch of neutral hydrogen gas at redshift $z$, 21~cm photons can either be absorbed from or emitted into this radiation. This is entirely dependent on the spin temperature ($T_S$) of the neutral hydrogen gas, which is defined by the number of hydrogen atoms in the triplet and singlet states. Usually, the primary source of background radiation is the CMB.  If the spin temperature is higher than the CMB temperature, the hydrogen will emit 21~cm radiation; conversely, if the spin temperature is lower, the hydrogen will absorb the 21~cm photons from the CMB. This 21~cm radiation, which gets redshifted into very low radio wavelengths by the time it reaches us, is usually measured relative to the CMB radiation field and expressed in terms of a differential brightness temperature, $\delta T_b$ (see \fig{fig:21cm} for an illustration).

The brightness temperature associated with the radiation emerging from neutral regions in the IGM is given by $T_{\rm b}(z) = T_{\rm CMB}~e^{-\tau_{21}}+T_S (1-e^{-\tau_{21}})$. Hence, the differential brightness temperature  observed at a frequency $\nu_{\rm obs}$ along a line of sight $\hat{\boldsymbol n}$ relative to the CMB can be expressed as,
\begin{align}
  \delta T_{\rm b}(\nu _{\rm obs}, \hat{\boldsymbol n})
  &= \frac{T_{\rm b}(z) - T_{\rm CMB} (z)}{1+z} 
  \nonumber \\
  &\simeq 27 \; {\rm mK}\; x_{\rm HI}(\boldsymbol {x},z) [1+\delta(\boldsymbol {x},z)] \left(1-\frac{T_{\rm CMB} (z)}{T_{\rm S}(\boldsymbol {x},z)}\right) \left(1 + \frac{1}{H(z)} \frac{\mathrm{d} v_p}{\mathrm{d} r}\right)^{-1}\left(\frac{1+z}{10}\right)^{1/2} \left(\frac{Y_H}{0.76}\right)\left(\frac{0.14}{\Omega_m h^2}\right)^{1/2} \left(\frac{\Omega_b h^2}{0.022}\right).
  \label{eq:deltaTB},
\end{align} 
where we have assumed that the optical depth $\tau_{21}$ of 21~cm radiation is much less than unity (the so-called ``optically thin'' limit). In the above expression, ${\mathbf{x}} = r_c \hat{\boldsymbol n} $ is the comoving distance along the line of sight to redshift $z$ ($\equiv 1420~{\rm MHz}/\nu_{\rm obs}-1$) while quantities $x_{\rm HI}(\mathbf{x},z)$ and $\delta(\mathbf{x},z)$ denote the neutral hydrogen fraction and the density contrast of baryonic gas respectively at point $\mathbf{x}$ and redshift $z$. The quantity $\mathrm{d} v_p/\mathrm{d} r$ denotes the gradient of the proper velocity projected to the line of sight. The spin temperature of the neutral hydrogen gas in the IGM is represented by the quantity  $T_S$, whereas $T_{\rm CMB}(z)$ = 2.73~$(1+z)$~K denotes the CMB brightness temperature at redshift $z$. 

By the time reionization begins, the IGM has already been significantly heated, primarily by X-ray photons (and to a lesser extent, by UV photons) from the first luminous sources. This causes the spin temperature of neutral hydrogen, which is coupled to the gas kinetic temperature, to be much higher than the CMB temperature and therefore, we observe the differential 21~cm signal in emission during the EoR. During this time, the evolution and fluctuations of the differential brightness temperature are essentially driven by the corresponding evolution and fluctuations in the gas density and ionization structure of the IGM and the peculiar velocities of the gas along the line of sight.

There are two main observational techniques for detecting the cosmological 21~cm signal from the Epoch of Reionization. The first method involves measuring the average intensity of 21~cm radiation across the entire sky as a function of frequency, known as the global 21~cm signal. The second approach focuses on detecting variations in the intensity of 21~cm radiation over a large region of the sky. This method not only aims to statistically detect the signal using summary statistics like the power spectrum or bispectrum but also strives to produce three-dimensional 21~cm images of the IGM during this epoch.
\begin{figure}[t]
\centering
\includegraphics[scale=0.5]{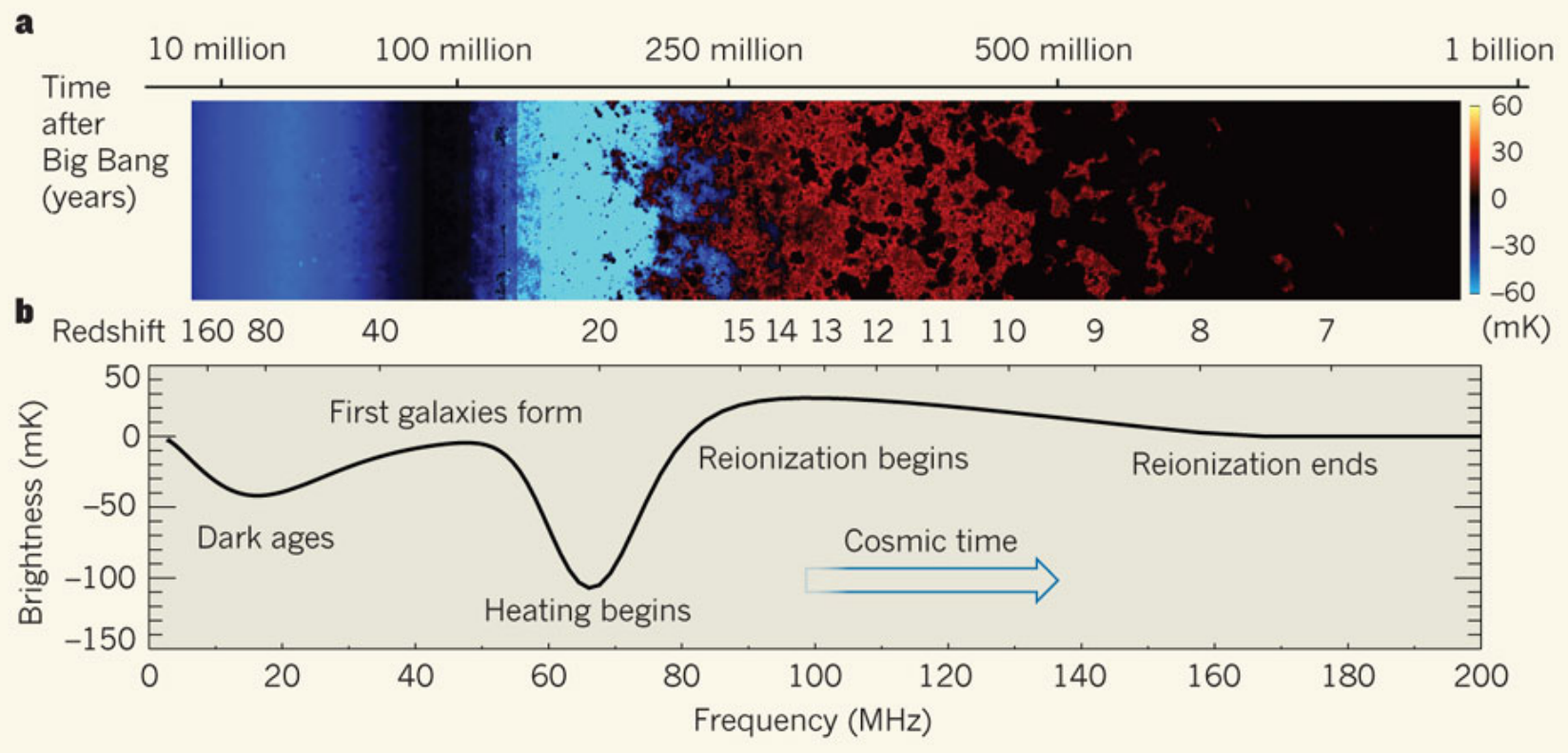}
\caption{The evolution of the brightness temperature of the 21~cm radiation, seen relative to the CMB temperature, as a function of time for a typical model of reionization and cosmic dawn. The upper panel shows the evolution of the strength of the spatial fluctuations in the 21~cm brightness temperature along a particular line of sight from a cosmological simulation while the lower panel shows the evolution of the sky-averaged 21~cm global signal, highlighting the key cosmic epochs beginning from the formation of the first stars and galaxies up to the completion of reionization. Image Credits: \cite{Pritchard2010Nature} }
\label{fig:21cm_evolution}
\end{figure}

The strength and shape of the global 21~cm signal carry important information about the timing of key cosmic events, such as the formation of the first stars, the onset of cosmic heating by these stars, and the beginning of reionization (\fig{fig:21cm_evolution} shows the evolution of the 21~cm signal in a typical model of cosmic dawn and reionization). However, the 21~cm fluctuations are expected to provide far more detailed insights, particularly regarding the abundance, nature, and distribution of the sources driving reionization, as well as the physical state of the IGM, that cannot be captured using only the globally averaged quantities. For example, the large-scale 21~cm power spectrum is expected to peak on spatial scales corresponding to the characteristic size of ionized regions at that epoch. However, a key difficulty in detecting the 21~cm line is that the cosmological signal is many orders of magnitude smaller than the astrophysical foreground signal arising from galactic and extra-galactic synchrotron and free-free emissions. There has been considerable work to explore ways to remove or avoid these foregrounds and extract the underlying signal. Another possible way to mitigate some part of the foreground contamination is to cross-correlate the 21~cm observations with other high-$z$ observables such as galaxies, which are not expected to exhibit any correlation with the 21~cm foreground signal. Such synergetic approaches are expected to not only provide unambiguous confirmation of any putative detection of the 21~cm signal based solely on statistical auto-correlation studies but also offer independent insights about the drivers and topology of reionization \citep{Lidz2009, Hutter2023}.

Currently, several interferometric experiments, including the Low Frequency Array (LOFAR) \citep{Gehlot2019_LOFAR, Mertens2020_LOFAR}, Murchison Widefield Array (MWA) \citep{Barry2019_MWA, Trott2020_MWA}, Precision Array for Probing the Epoch of Reionization (PAPER) \citep{Parsons2010_PAPER, Abdurashidova2022_HERA}, and Hydrogen Epoch of Reionization Array (HERA) \citep{Kolopanis2019_PAPER}, are focused on detecting the power spectrum of 21~cm fluctuations. These efforts are complemented by global 21~cm experiments, such as the Shaped Antenna measurement of the background RAdio Spectrum (SARAS) \citep{Singh2022_SARAS}, Experiment to Detect the Global EoR Signature (EDGES) \citep{Bowman2018_EDGES}, and Radio Experiment for the Analysis of Cosmic Hydrogen (REACH) \citep{Acedo2022_REACH}, which aim to measure the 21~cm global signal over a wide range of epochs spanning both the Epoch of Reionization and Cosmic Dawn. In the near future, upcoming telescopes like the Square Kilometer Array (SKA) \citep{Koopmans2015_SKA}, with their extremely high resolution and sensitivity, will most likely revolutionize the field of 21~cm cosmology by providing us with a comprehensive 3D tomographic view of the IGM  during reionization.

\subsection{Fast Radio Bursts}

Another promising probe for studying reionization that has gained significant attention in recent years is the use of fast radio bursts (FRBs). FRBs are extremely bright ($\sim$ 10$^{-26}$ watts per square metre per hertz) and short-duration ($\sim$ millisecond) pulses of radio emission coming from extragalactic distances. As these pulses travel towards us, they get dispersed due to their interaction with free electrons in the intervening medium. The amount of dispersion suffered, quantified by the Dispersion Measure (DM), directly depends on the free electron content along its path. For a cosmological FRB, the dominant contribution to its total observed DM is expected to arise from the intergalactic medium ($\mathrm{DM}_{\mathrm{IGM}}$), with additional contributions resulting from the host galaxy and the Milky Way. As reionization will only influence the IGM contribution ($\mathrm{DM}_{\mathrm{IGM}}$) to the total ``observed'' DM, it is essential to accurately characterize the other two contributory sources when using FRBs to study the EoR.

Therefore, an ensemble of localized FRBs spread out over the reionization epoch, having precise intergalactic DM measurements and independent redshift estimates from optical observations, can in principle be used to fetch information about the reionization history with the help of simplistic estimators such as the sky-averaged intergalactic DM (i.e., $\langle \mathrm{DM}_{\mathrm{IGM}} \rangle (z)$) or the redshift derivative of the sky-averaged intergalactic DM (i.e., $\mathrm{d}\langle {\rm DM_{\rm IGM}}\rangle /\mathrm{d}z$) \citep{Beniamini2021, Heimersheim2022, Shaw2024, Maity2024}. Interestingly, some studies (e.g., \citep{Fialkov2016}) have argued that a series of measurements of $\langle \mathrm{DM}_{\mathrm{IGM}} \rangle$ at various cosmic epochs up to some given redshift $z$ can be used to determine the mean integrated Thomson scattering optical depth out to the same redshift, $\tau(z)$ since both these quantities are integrals of the electron number density with slightly different redshift-weight factors. Building on this idea, they propose that integrating this $\tau - \mathrm{DM}$ relation over the entire range of DMs up to the maximum observed $\mathrm{DM}_{\mathrm{IGM}}$ in a sample of localized FRBs, spanning the period from the onset of reionization to the present day, could provide independent constraints on the total electron scattering optical depth ($\tau_{\mathrm{el}}$) of CMB photons \citep{Fialkov2016}.

\section{Conclusion}
%A concluding paragraph summing up your main points in the chapter 
Studying the epoch of reionization is crucial to understanding the early Universe and the formation of the first luminous objects, such as stars, galaxies, and quasars. This period, which began a few hundred million years after the Big Bang, marks a transformative phase in cosmic history when the neutral hydrogen in the intergalactic medium (IGM) was ionized by the ultraviolet radiation emitted by these early sources. By investigating reionization, it is possible to gain insight into the formation and evolution of the first galaxies and black holes, as well as the feedback mechanisms that influenced subsequent galaxy evolution. Additionally, reionization provides essential information about the large-scale structure of the Universe, as the ionization bubbles created by galaxies trace the underlying cosmic web of matter distribution. Understanding reionization also helps interpret observations from contemporary and future telescopes, such as the James Webb Space Telescope (JWST) and the upcoming Square Kilometre Array (SKA), offering a clearer picture of the Universe’s transition from the ``Dark Ages'' to the complex structures we observe today.

\begin{ack}[Acknowledgments]

The authors acknowledge support from the Department of Atomic Energy, Government of India, under project no. 12-R\&D-TFR-5.02-0700.

\end{ack}

\seealso{ \cite{Barkana2001_Review, Choudhury2009_Review, Loeb2013_Book, Mesinger2016ASSL, Dayal2018_Review, Choudhury2022_Review, Gnedin2022}}

\bibliographystyle{Harvard}
\bibliography{reference}

\end{document}